# Lieb lattices formed by real atoms on Ag(111) and their lattice constant dependent electronic properties

Xiaoxia Li[1]†, Qili Li[1]†, Tongzhou Ji[1], Ruige Yan[1], Wenlin Fan[1], Bingfeng Miao[1,2],

Liang Sun[1,2], Gong Chen[1,2], Weiyi Zhang[1,2], Haifeng Ding[1,2,*]

[1]National Laboratory of Solid State Microstructures and Department of Physics,

Nanjing University, Nanjing 210093, China.

[2]Collaborative Innovation Center of Advanced Microstructures, Nanjing 210093, China.

*Corresponding author.

†These authors contribute equally to this work.

Email: hfding@nju.edu.cn

## Abstract

Scanning tunneling microscopy is a powerful tool to build artificial atomic structures even not exist in nature but possess exotic properties. We here constructed Lieb lattices with different lattice constants by real atoms, i.e., Fe atoms on Ag(111) and probed their electronic properties. We find a surprising long-range effective electron wavefunction overlap between Fe adatoms as it exhibits a $\frac{1}{r^2}$-dependence with the interatomic distance $r$ instead of the theoretically predicted exponential one. Combining control experiments, tight-binding and Green's function calculations, we attribute the observed long-range overlap to be enabled by the surface state. Our findings not only enrich the understanding of the electron wavefunction overlap, but also provide a convenient platform to design and explore the artificial structures and future devices with real atoms.

Artificial atomic structures built by scanning tunneling microscopy (STM) are a fertile playground to investigate fundamentals and explore potential applications. For instance, logic gates are realized by CO molecule cascades on Cu(111) [1] and Fe elliptical quantum corrals on Ag(111) [2]. Quantum holographic encoding of CO on Cu(111) achieves information densities in excess of 20 bits/nm$^2$ [3]. A kilobyte rewritable atomic memory is presented for Cl vacancies on Cu(100) [4]. Fundamentally, Au atomic chains on NiAl(110) demonstrate the development of one-dimensional band structure [5]. Spin-spin interactions are probed for Mn chains on CuN on Cu(100) [6]. Majorana bound states are observed at the ends of chains for Fe on Pb(110) [7-9] and Re(0001) [10]. Besides, two-dimensional structures are also widely studied for quantum size effect in nanocorrals [11-13], Dirac fermions in molecular graphene of CO on Cu(111) [14], quasi-crystals in Penrose tiling of CO on Cu(111) [15], fractals in Sierpiński triangle CO on Cu(111) [16], etc.

Lieb lattice [17], a two dimensional square lattice consisting of an atom at the corner and two atoms at the middle of each edge, attracts great interests due to its exotic electronic band structure and predicted unusual properties in ferromagnetism [17-20], superconductivity [21-23] and topological states [24-26], despite it does not exist in nature. Recently, Lieb lattices have been experimentally realized in optical [27-30] and electronic systems with artificial objects [31-34]. As aforementioned, STM naturally has the advantage to study this structure. Indeed, Lieb Lattices have been successfully assembled and investigated with Cl vacancies on Cu(100) [32] and artificial atoms formed by the quantum confinement of the surface state through the CO molecules on Cu(111) [31]. We note that in the study of the Lieb lattices realized by the Cl vacancies and quantum states confined by CO molecules, artificial objects are used to mimic the real atoms. It would be more desirable to construct the Lieb lattices with the real atoms and investigate their properties.

Here, we utilize atomic manipulation [35] to construct Lieb lattices with the real atoms, i.e., Fe adatoms on Ag(111) instead of the anti-Lieb lattices formed by Cl or CO molecules. We find the typical features of Lieb lattice by measuring the spectroscopy and differential conductance map. We further performed systematically lattice constant-dependent studies by tuning the interatomic distance $r$ through atomic manipulation. Note that the high internal pressure in solid makes it nearly impossible to vary $r$ with large amplitude, which is imperative for the $r$-dependence studies. In comparison with tight-binding modelling, we find that the effective overlap energy $t$ shows a $\frac{1}{r^2}$-dependence, in sharp contrast with the theoretically predicted exponential one for the direct overlap between two *s*-states [36]. Combining control experiments and Green's function based calculations [37, 38], we identify that the observed long-range features of Lieb lattices is an indirect interaction between Fe atoms mediated by the surface state.

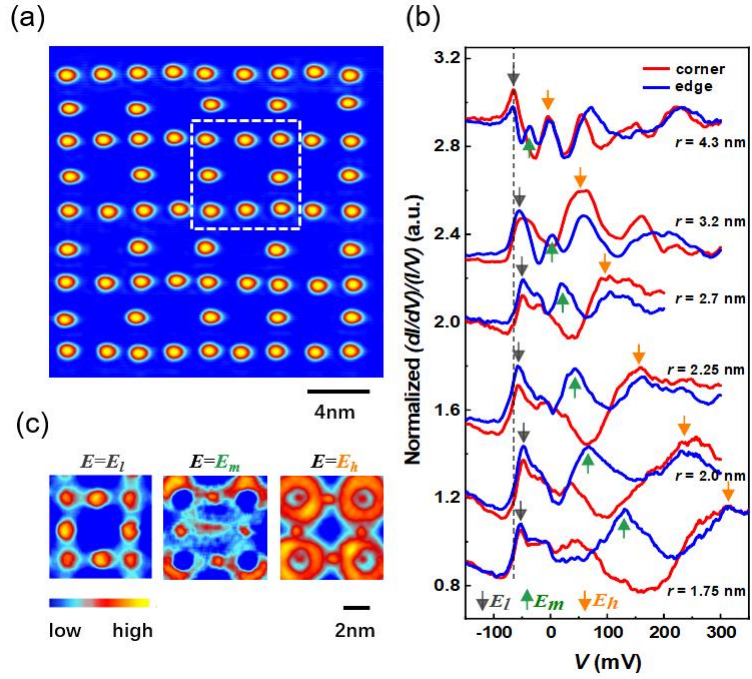

**Fig. 1.** (a) Typical topographic image of a 4×4 Lieb lattice ($r$ = 2.25 nm) constructed with Fe adatoms on Ag(111). (b) Normalized (*dI/dV*)/(*I/V*) spectra at corner sites (red) and edge sites (blue) with different $r$ (curves are shifted for clarity). Dash line indicates the surface state onset energy of Ag(111). Orange, green and gray arrows mark the high-, middle- and low-energy peaks, respectively. (c) Typical *dI/dV* maps obtained at the square area marked in (a) at the low, middle and high energy peaks.

The experiments were performed in a low-temperature STM system with the base pressure of $2\times10^{-11}$ mbar. The Ag(111) single crystals were prepared with cycles of Ar+ sputtering (1.5kV) and annealing (~580 °C). High-purity Fe was deposited onto the Ag(111) surfaces with a typical rate of 0.002 monolayer equivalent per minute by means of electron beam evaporation at ~6 K. The measurements and atomic manipulations are performed with the W tips at ~4.7 K. We constructed a series of Lieb lattices by laterally manipulating Fe atoms on the Ag(111) surface [2, 35]. Figure 1(a) shows a typical topographic image of an assembled 4×4 Lieb lattice ($r$ = 2.25 nm). To investigate its electronic properties, we acquire both the tunneling current $I$ and the differential conductance ($dI/dV$) as the function of the bias voltage $V$ on top of the Fe adatoms at the corner- and edge-sites (Fig. S5(a)), as well as on an isolated single adatom via the lock-in technique with a modulation of the sample voltage of 4 mV at a frequency of 6.3 kHz (Fig. S4(b)). The tip is stabilized at 50 mV and 1 nA prior the $dI/dV$ measurements. To exclude the tip effect, we normalized the raw data through dividing the $(dI/dV)/(I/V)$ curves of the Fe adatoms in the lattices by the one obtained on top the isolated Fe adatom with the same tip. The normalization also minimizes the influence of the tunneling matrices.

As shown in Fig. 1(b), at the edge sites (blue curves), the normalized spectra show three pronounced sets of peaks: the low energy peaks (gray arrows) located close to the surface state onset energy (gray dashed line), −65 mV [39], the middle energy peaks (indicated by green arrows) and the high energy peaks (marked by orange arrows). Interestingly, at the corner sites (red curves), they exhibit only two sets of peaks that coincide with the peaks near the onset energy of the surface state and the high energy peaks obtained at the edge sites. When the value of $r$ increases from 1.75 to 4.8 nm, both sets of the middle and high energy peaks shift to lower energy while the low energy one remains nearly unchanged. In addition, the middle energy peak is located approximately at the energy in the middle of the high and low energy peaks. We note that there are also other peaks at the energies above these three peaks. They may originate from the scattering of the surface state since Fe adatoms also serve as the scattering centers for the surface state, as will be confirmed by the Green's function calculation in supplementary materials. The measured $dI/dV$ maps for $r$ = 2.25 nm at three energies are shown in Fig. 1(c), which indicates that the electrons are localized at edges sites at the energy of $E_m$, while they distribute both at the corner and edge sites at both energies of $E_l$ and $E_h$. The features in the middle of the Lieb lattice at $E_m$ and the ring-shaped features at $E_h$ in Fig. 1(c) are the interference pattern caused by the scattering of the surface state as will be further discussed in supplementary material. We note that the $dI/dV$ maps were obtained with a constant height mode. Due to the chemical difference, the actual height between the tip and the surface might be different for the tip placed on top of Ag and Fe. Here, we mainly focus on the spectra on top of the Fe adatoms.

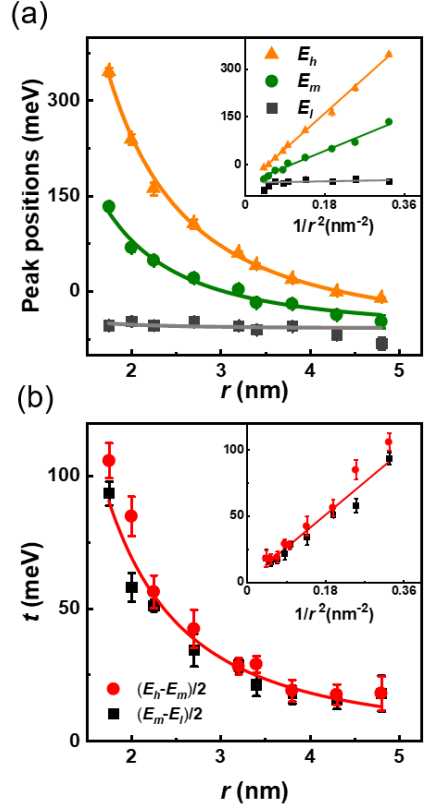

**Fig. 2**. (a) Peak positions of the normalized differential conductance spectra as a function of $r$ and $1/r^2$ (inset). (b) The comparison of $(E_h - E_m)/2$ and $(E_m - E_l)/2$ as a function of $r$ and $1/r^2$ (inset). The curves outside (inside) the insets are the fittings with the inverse parabolic (linear) function.

To analyze the variation of the peak position with the atomic separation $r$, we plot $E_l$, $E_m$, and $E_h$ as a function of $r$ in Fig. 2(a) and found that both data sets can be fitted with $E = E_0 + \dfrac{C}{r^2}$, where the value of $E_0$ is close to the onset energy of the surface state (Table S1). As shown in Fig. 2(b), we also obtained the $r$-dependent values of $(E_h - E_m)/2$ and $(E_m - E_l)/2$. Interestingly, we find that they are very similar within our experimental error margin. As the peak positions $E_l$, $E_m$ and $E_h$ show the $\frac{1}{r^2}$-dependence, we fitted them with $\Delta E = \Delta E_0 + \frac{C_\Delta}{r^2}$. As shown in Fig. 2(b), the fitted curve with fitting parameters $\Delta E_0 = 1.0 \pm 2.2$ meV and $C_\Delta = 273.9 \pm 14.5$ meV·nm$^2$ reproduces the experimental data well, which evidences that the half of the peak interval is $\frac{1}{r^2}$-dependent. For clarity, we also plot them as a function of $1/r^2$ and show their linear relationship in the insets of Fig. 2.

To understand the observed electronic properties of the lattices, we first calculated the band structure and LDOS by tight-binding method by only considering the *s*-orbitals of the real atoms, which means no substrate is considered. We set the onsite energy to be zero for both corner and edge site and the nearest neighboring overlap energy to *t*. As shown in Fig. 3(a), the band structure features two Dirac bands and a flat band [40-42]. Correspondingly, the LDOS curve shows two peaks at the corner site and three peaks at the edge site with the middle peak located at the center of the other two peaks. Moreover, the calculated LDOS maps show that the electrons are mainly localized at edge sites at the middle energy, while the electrons are distributed at both corner and edge sites at low and high energy. We note that, in the above tight-binding calculations, we only consider the overlap between *s*-orbitals. The Hamiltonian, however, should be also valid when the nearest neighboring sites have relatively localized states which overlap with an effective overlap energy. For Fe adatom on Ag(111), the 4*s* state has a strong coupling with the surface state, resulting relatively localized states. It can be anticipated that the band structure formed by these localized states are similar as the one formed by the *s*-orbitals, except that the onsite energy and overlap energy are different. Therefore, the tight-binding calculated results essentially reproduce the features on normalized (*dI/dV*)/(*I/V*) curves and the *dI/dV* maps (Figs. 1(b) and 1(c)) observed in our experiments. The experimentally obtained $E_l$, $E_m$ and $E_h$ are thus attributed to the characteristic peaks of the Lieb lattice. Notably, as shown in Fig. 3(a), the peak interval is 2*t* (4*t*) at the edge (corner) site, i.e., the overlap energy *t* is $(E_h - E_m)/2$ or $(E_m - E_l)/2$. Thus, the peak intervals in the experimental *dI/dV* curves correspond to twice of the overlap energy and it is $\frac{1}{r^2}$-dependent. We note that, in the tight-binding calculation shown in Fig. 3, we only consider the nearest neighbor hopping, the band structure is symmetric around the onsite energy. When the second nearest neighbor hopping is included, the band structure is no longer symmetric.

The observed $\frac{1}{r^2}$-dependent overlap energy in our experiments is, however, in sharp contrast with the exponential dependence predicted in the textbook [36]. The prediction was made for the direct overlap between the *s*-states of two hydrogen atoms in free space. The direct overlap energy can be calculated to be ~10 meV when their separation is 0.55 nm. In our experiments, the measured overlap energy is ~17 meV even when *r* is 4.8 nm. Therefore, the long-range overlap energy we observed on Ag(111) is not the direct overlap between Fe atoms. We attribute the difference between the theoretical prediction and our experiments to the atomic environment as the prediction was made for atoms in free space while our experiments are performed with adatoms placed on top of a Ag(111) surface, the latter enables a surface state around the Fermi energy. Moreover, we performed similar measurements on a Ag(100) substrate, which has no surface state near the Fermi energy. Interestingly, the spectra of the Fe Lieb lattices on Ag(100) do not show any apparent feature of the Lieb lattice (Fig. S10). Instead, the spectra obtained either at the corner- or edge-sites are very similar to the spectrum of the isolated single adatom even when *r* approaches only ~1 nm. The significantly different *r*-dependence for Fe adatoms on both substrates highlights the crucial role of the surface state on the observed long-range overlap on Ag(111) surface. As previous studies, the spectrum of the isolated Fe adatom (Fig. S4(b)) shows that the 4*s* state of Fe has a strong overlap with the surface state, resulting relatively localized states [43, 44]. It can be anticipated that the band structure formed by these localized states are similar as the one formed by the *s*-orbitals, except that the onsite energy and overlap energy are different. Thus, the electronic features of Lieb lattice are observed at a large atomic separation. In fact, this is not too surprising as in the exploration of the quantum mirage, the Kondo resonance peak of Co adatom located at one focus point of elliptical corral can be projected to the other focus point around 10 nm apart on Cu(111) via the assistance of the surface state [12]. In a very similar study, the inversion state of an Fe adatom can also be transferred to another location more than 10 nm apart in Ag(111) through the surface state [2].

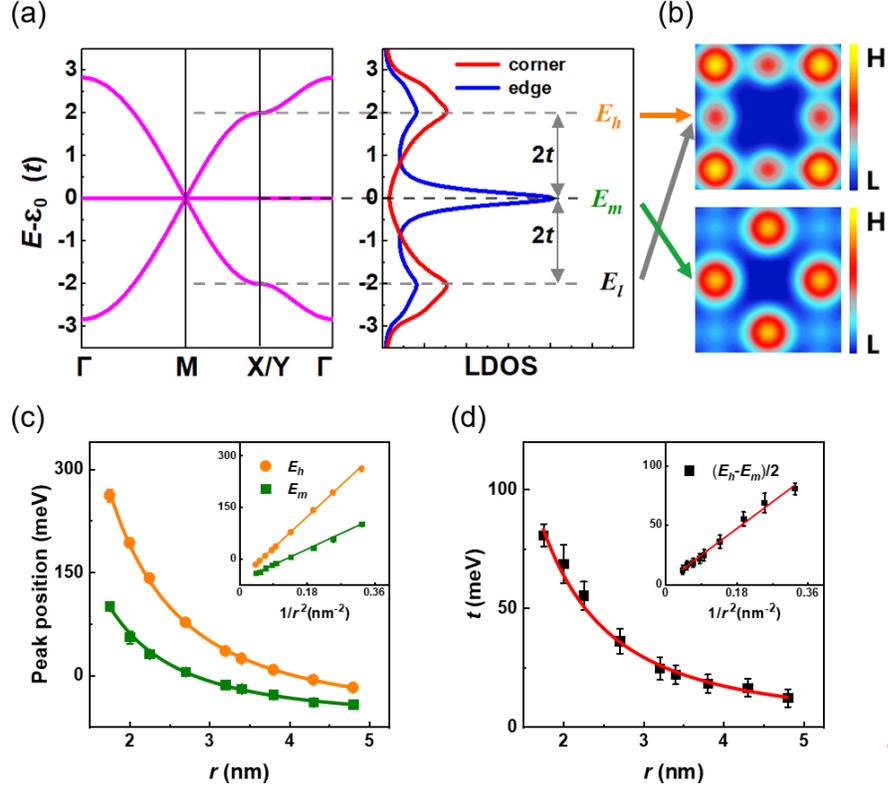

**Fig. 3.** (a) Calculated band structure of a Lieb lattice and the corresponding LDOS at corner (red) and edge (blue) sites with the effective nearest neighboring overlap energy $t$ through the tight-binding method. Note: The energy differences between $E_h$, $E_m$ and $E_l$ are $2t$ as marked. (b) LDOS maps at $E_h$, $E_m$ and $E_l$, respectively. (c) Calculated peak positions of a $4\times4$ Lieb lattice on Ag(111) via the Green's function method as the function of $r$ and $1/r^2$ (inset). (d) Corresponding value of $(E_h - E_m)/2$. The curves outside (inside) the insets in (c) and (d) are fittings with the inverse parabolic (linear) functions.

To further verify the important role of surface state, we performed Green's function-based calculations for Fe Lieb lattices built on Ag(111). The calculated LDOS curves exhibit close similarity with the experimental *dI/dV* (Fig. S5(c)). Note that we mainly consider surface state in the Green's function calculations, which are valid only above the onset of the surface state. In addition, the experimentally obtained peak $E_l$ is very close to the onset energy, which corresponds to small wavevector and large wavelength. Thus $E_l$ might be influenced by the onset of the surface state and not accurately determined. Therefore, we focus our discussion on $E_m$ and $E_h$. As shown in Fig. 3(c), we plot $E_m$ and $E_h$ as a function of $r$ and they can be fitted well with $E = E_0 + \frac{C}{r^2}$. Furthermore, the value of $(E_h - E_m)/2$ also follows the $\frac{1}{r^2}$-dependence. The fitting parameter $C_\Delta = 249.3 \pm 8.1$ meV·nm² agrees well with our experimental value of $273.9 \pm 14.5$ meV·nm². The insets in Figs. 3(c) and 3(d) show the linear dependences of the peak positions and $(E_h - E_m)/2$ as a function of $1/r^2$, respectively. We note that, in the Green's function calculation, we not only consider the scattering of the surface state by the Fe adatom but also the coupling of Fe 4*s* state and the surface state of Ag(111) with the empirical description of the inversion effect. For more accurate analysis, first-principle calculations may be needed.

In summary, we successfully constructed Fe Lieb lattice on Ag(111) surface by real atoms and tuned its electronic properties by varying its atomic separation. We found that the effective overlap energy between lattice site exhibits a $\frac{1}{r^2}$-dependence. Combining the control experiments on Ag(100), tight-binding and Green's function calculations, we attribute the long-range overlap energy to the effective overlap of Fe adatom mediated by the surface state. Note that in order to observe this long-range effect between lattice adatoms, the hybridization between adatoms and surface state needs to be strong enough. It is known that Co, Ag, Cu, Mn adatoms also show strong hybridization on Ag(111), Cu(111) or Au(111) [37, 45-49] besides Fe adatoms. Thus, our experiments provide a convenient platform to design and explore the artificial structures and devices with various materials combination and exotic properties, such as flat-band and topological band etc. We note that similar approaches were successfully used to realize atomic systems but with direct bonding [50, 51] as well as the flatband system, molecule graphene, quasi-crystal, etc., but with artificial atoms [14, 15, 52].

This work was supported by the National Key R&D Program of China (grants no. 2017YFA0303202, and no. 2018YFA0306004); the National Natural Science Foundation of China (grants no. 11974165, no. 51971110, and no. 11734006); the China Postdoctoral Science Foundation (grant no. 2019M651766); and the Natural Science Foundation of Jiangsu Province (grant no. BK20190057).